\documentclass[5p,sort&compress]{elsarticle} 

\RequirePackage{lineno}
\usepackage{units}
\usepackage{xspace}
\usepackage{mathrsfs}
\usepackage{units}
\usepackage{amssymb}
\usepackage{placeins}
\usepackage{subfigure}
\usepackage{booktabs}
\usepackage{epsfig}

\usepackage{preprintcover} 
\PreprintCoverPaperTitle{Search for direct production of charginos and neutralinos in events with three leptons and missing transverse momentum in $\sqrt{s}\,=\,$7$\,$TeV $pp$ collisions with the ATLAS detector}  
\PreprintIdNumber{CERN-PH-EP-2012-217}  
\PreprintCoverAbstract{A search for the direct production of charginos and neutralinos in final states with three electrons or muons and  missing transverse momentum is presented. The analysis is based on 4.7$\,$fb$^{-1}$ of $\sqrt{s}\,=\,$7$\,$TeV proton-proton collision data delivered by the Large Hadron Collider and recorded with the ATLAS detector. Observations are consistent with Standard Model expectations in three signal regions that are either depleted or enriched in $Z$-boson decays. Upper limits at 95\% confidence level are set in R-parity conserving phenomenological minimal supersymmetric models and in simplified models, significantly extending previous results.   } 
\PreprintJournalName{Physics Letters B}

\newcommand{\Alpgen}{{\tt ALPGEN}\xspace}
\newcommand{\Jimmy}{{\tt JIMMY}\xspace}
\newcommand{\Mcatnlo}{{\tt MC@NLO}\xspace}
\newcommand{\Sherpa}{{\tt SHERPA}\xspace}
\newcommand{\Herwig}{{\tt HERWIG}\xspace}
\newcommand{\Herwigpp}{{\tt Herwig++}\xspace}

\newcommand{\Prospino}{{\tt PROSPINO}\xspace}
\newcommand{\Madgraph}{{\tt MadGraph}\xspace}

\newcommand{\CTEQSixL}{{\tt CTEQ6L1}\xspace}
\newcommand{\CTEQSix}{{\tt CTEQ6.6}\xspace}
\newcommand{\CTTen}{{\tt CT10}\xspace}
\newcommand{\PYTHIA}{{\tt PYTHIA}\xspace}
\newcommand{\GEANT}{{\tt GEANT4}\xspace}
\newcommand{\MSTW}{{\tt MSTW}\xspace}

\newcommand{\MCFM}{{\tt MCFM}\xspace}
\newcommand{\fewz}{{\tt FEWZ}\xspace}
\newcommand{\MSTWloc}{{\tt MRTSmcal}\xspace}
\newcommand{\antibar}[1]{\ensuremath{#1\bar{#1}}}
\newcommand{\ttbar}{\antibar{t}\xspace}

\newcommand{\ninoone}{\ensuremath{\tilde{\chi}_1^0}\xspace}
\newcommand{\ninotwo}{\ensuremath{\tilde{\chi}_2^0}\xspace}
\newcommand{\chinoone}{\ensuremath{\tilde{\chi}_1^\pm}\xspace}

\newcommand{\met}{\ensuremath{E_{\mathrm{T}}^{\mathrm{miss}}}\xspace}

\def\ninoi{\ensuremath{\mathchoice
      {\displaystyle\raise.4ex\hbox{$\displaystyle\tilde\chi^0_j$}}%
         {\textstyle\raise.4ex\hbox{$\textstyle\tilde\chi^0_j$}}%
       {\scriptstyle\raise.3ex\hbox{$\scriptstyle\tilde\chi^0_j$}}%
 {\scriptscriptstyle\raise.3ex\hbox{$\scriptscriptstyle\tilde\chi^0_j$}}}}

\def\chinojpm{\ensuremath{\mathchoice%
      {\displaystyle\raise.4ex\hbox{$\displaystyle\tilde\chi^\pm_i$}}%
         {\textstyle\raise.4ex\hbox{$\textstyle\tilde\chi^\pm_i$}}%
       {\scriptstyle\raise.3ex\hbox{$\scriptstyle\tilde\chi^\pm_i$}}%
{\scriptscriptstyle\raise.3ex\hbox{$\scriptscriptstyle\tilde\chi^\pm_1$}}}}

\journal{Physics Letters B}

\begin{document}

\begin{frontmatter}

 \title{Search for direct production of charginos and neutralinos in events with three leptons and missing transverse momentum in $\sqrt{s}\,=\,$7$\,$TeV $pp$ collisions with the ATLAS detector}

\author{The ATLAS Collaboration}

\begin{abstract} 
A search for the direct production of charginos and neutralinos in final states with three electrons or muons and  missing transverse momentum is presented. The analysis is based on 4.7$\,$fb$^{-1}$ of $\sqrt{s}\,=\,$7$\,$TeV proton-proton collision data delivered by the Large Hadron Collider and recorded with the ATLAS detector. Observations are consistent with Standard Model expectations in three signal regions that are either depleted or enriched in $Z$-boson decays. Upper limits at 95\% confidence level are set in R-parity conserving phenomenological minimal supersymmetric models and in simplified models, significantly extending previous results.   
\end{abstract}

\end{frontmatter}

\section{Introduction}

Supersymmetry (SUSY)~\cite{Miyazawa:1966,Ramond:1971gb,Golfand:1971iw,Neveu:1971rx,Neveu:1971iv,Gervais:1971ji,Volkov:1973ix,Wess:1973kz,Wess:1974tw} postulates the existence of SUSY particles, or ``sparticles", with spin differing by one-half unit with respect to that of their Standard Model (SM) partner. If R-parity~\cite{Fayet:1976et,Fayet:1977yc,Farrar:1978xj,Fayet:1979sa,Dimopoulos:1981zb} is conserved, the lightest SUSY particle (LSP)  is stable and 
sparticles can only be pair-produced and decay into final states with SM particles and  LSPs. 
 Charginos ($\chinojpm$, $i=1,2$) and neutralinos ($\ninoi$, $j=1,2,3,4$) are the mass eigenstates formed from the linear superposition of the SUSY partners of the Higgs and electroweak gauge bosons. These are the Higgsinos, and the winos, zino, and bino, collectively known as gauginos. Naturalness requires $\chinojpm$ and $\ninoi$ (and third-generation sparticles) to have masses in the hundreds of GeV range~\cite{Barbieri1987fn,deCarlos1993yy}. In scenarios where squark and gluino masses are larger than a few TeV, the direct production of gauginos may be the dominant SUSY process at the Large Hadron Collider (LHC). Charginos can decay into leptonic final states via sneutrinos ($\tilde{\nu}\ell$), sleptons ($\tilde{\ell}\nu$) or $W$ bosons ($W\ninoone$), while unstable neutralinos  can decay via  sleptons ($\ell\tilde{\ell}$) or $Z$ bosons ($Z\,\ninoone$). 

This Letter presents a search with the ATLAS detector for the direct production of charginos and neutralinos decaying to a final state with three leptons (electrons or muons) and missing transverse momentum, the latter originating from the two undetected LSPs and the neutrinos. The analysis is based on 4.7$\,$fb$^{-1}$ of proton-proton collision data delivered by the LHC at a centre-of-mass energy $\sqrt{s}\,$$=\,$7$\,$TeV between March and October 2011. 
The search described here significantly extends the current mass limits on charginos and neutralinos set by ATLAS~\cite{ATLAS2lep2011,ATLASRel163LSUSY}. 
Similar searches have been conducted at the Tevatron~\cite{D0-2009,CDF-2008}  and LEP~\cite{PDG},  where a model-independent lower limit of 103.5$\,$GeV was set at 95\% confidence level (CL) on the mass of promptly decaying charginos.

\section{Detector Description}

ATLAS~\cite{atlas-det} is a multipurpose particle detector with forward-backward symmetric cylindrical geometry. It includes an inner tracker (ID)  immersed in a 2$\,$T magnetic field providing precision tracking of charged particles for pseudorapidities $|\eta|\,$$<\,$2.5~\footnote{ATLAS uses a right-handed coordinate
  system with its origin at the nominal interaction point (IP) in the
  centre of the detector and the $z$-axis along the beam pipe. The
  $x$-axis points from the IP to the centre of the LHC ring, and the
  $y$-axis points upward. Cylindrical coordinates $(r,\phi)$ are used
  in the transverse plane, $\phi$ being the azimuthal angle around the
  beam pipe. The pseudorapidity is defined in terms of the polar angle
  $\theta$ as $\eta=-\ln\tan(\theta/2)$.}.   
Calorimeter systems with either liquid argon or scintillating tiles as the active media provide energy measurements over the range $|\eta|\,$$<\,$4.9. The muon detectors are positioned outside the calorimeters and are contained in an air-core toroidal magnetic field produced by superconducting magnets with field integrals varying from 1$\,$T$\cdot$m to 8$\,$T$\cdot$m. They provide trigger and high-precision tracking capabilities for $|\eta|\,$$<\,$2.4 and $|\eta|\,$$<\,$2.7, respectively.

\section{New Physics Scenarios}
In this analysis, results are interpreted in the phenomenological minimal supersymmetric SM (pMSSM \cite{pMSSM}) and in simplified models~\cite{simplModels1}.  

In the pMSSM the mixing for the $\chinojpm$ and $\ninoi$ 
depends on the gaugino masses $M_1$ and $M_2$, the Higgs mass parameter $\mu$, and $\tan\beta$, the ratio of the expectation values of the two Higgs doublets. 
The dominant mode for gaugino production leading to three-lepton final states is $\chinoone\ninotwo$ production via the $s$-channel exchange of a virtual gauge boson. Other  $\chinojpm\ninoi$  processes contribute a maximum of 20\% to three-lepton final states depending on the values of the mass parameters.
The right-handed sleptons (including third-generation sleptons) are assumed to be degenerate and have a mass $m_{\tilde{\ell}_R}\,$$=\,$$(m_{\tilde{\chi}_2^{0}} + m_{\tilde{\chi}_1^{0}} )/2$, set via the right-handed SUSY-breaking slepton mass parameter at the electroweak scale. 
In these scenarios, decays to sleptons are favoured. The parameter $\tan\beta$ is set to 6, yielding comparable branching ratios into each slepton generation.  The masses of the gluinos, squarks and left-handed sleptons are chosen to be larger than 2$\,$TeV. 
In order to achieve maximum mixing in the top squark sector the  corresponding trilinear couplings are set to non-zero values, while  all other trilineal couplings are set to zero.

In the  simplified models considered, the masses of the relevant particles ($\chinoone$, $\ninotwo$, $\tilde{\chi}_1^{0}$, $\tilde{\nu}$, $\tilde{\ell}_L$) are the only free parameters. The charginos and heavy neutralinos are set to be wino-like and mass degenerate, and the lightest neutralino is set to be bino-like. Two different scenarios are considered. In the first case, the $\chinoone$ and $\ninotwo$ are pair-produced  and decay via left-handed sleptons, including staus, and sneutrinos of mass $m_{\tilde{\nu}}=m_{\tilde{\ell}_L}\,$$=\,$$(m_{\tilde{\chi}_1^{0}} + m_{\tilde{\chi}_1^{\pm}} )/2$ with a branching ratio of 50\% each. In the second scenario, the $\chinoone$ and $\ninotwo$ decay via $W$ and $Z$ bosons.

\section{Monte Carlo simulation}
Several Monte Carlo (MC) generators are used to simulate SM processes and new physics signals relevant for this analysis. \Sherpa~\cite{SherpaMC} is used to simulate diboson processes $WZ$ and $ZZ$. These include all diagrams leading to three leptons and one neutrino, and to four leptons, respectively, including internal conversions (virtual photons converting into lepton pairs).      
\Herwig~\cite{herwig} is used for $WW$, while \Madgraph~\cite{madgraph} is used for the $t\bar{t}W$,  $t\bar{t}WW$, $t\bar{t}Z$, $W\gamma$ and $Z\gamma$ processes. \Mcatnlo~\cite{mcatnlo} is chosen for the simulation of single- and pair-production of top quarks, and \Alpgen~\cite{alpgen} is used to simulate $W/Z$+jets. 
Expected diboson yields are normalised using next-to-leading-order (NLO) QCD predictions obtained with \MCFM~\cite{mcfm1,mcfm2}. The top-quark pair-production contribution is normalised to approximate next-to-next-to-leading-order calculations (NNLO)~\cite{hathor} and the $t\bar{t}W(W)/Z$ contributions are normalised to NLO~\cite{ttZ, ttW}. 
The $W\gamma$ and $Z\gamma$  yields are normalised to be consistent with the ATLAS cross-section measurement~\cite{wgamma}. 
The QCD NNLO \fewz~\cite{fewz1,fewz2} cross-sections are used for normalisation of the inclusive $W$+light-flavour jets and $Z$+light-flavour jets. The ratio of the NNLO to LO cross-section is used to rescale the $W$+heavy-flavour jets and  $Z$+heavy-flavour jets LO cross-sections.

The choice of the parton distribution functions (PDFs) depends on the generator. The \CTEQSixL~\cite{CTEQ6L1} PDFs are used with \Madgraph\ and \Alpgen, and the \CTTen~\cite{CT10pdf} PDFs with \Mcatnlo and \Sherpa. The \MSTWloc PDF set~\cite{pdf:mrst} is used for \Herwig. 

The pMSSM samples are produced with \Herwig\ and the simplified model samples with \Herwigpp~\cite{herwigplusplus}. The yields of the SUSY samples are normalised to the NLO cross-sections obtained from \Prospino~\cite{Beenakker:1996ch} using the PDF set \CTEQSix with the renormalisation/factorisation scales set to the average of the relevant gaugino masses.

Fragmentation and hadronisation for the \Alpgen and \Mcatnlo (\Madgraph) samples are performed with \Herwig (\PYTHIA~\cite{pythia}), while for \Sherpa, these are performed internally.   \Jimmy~\cite{jimmy} is interfaced to \Herwig for simulating the underlying event.  
For all MC samples, the propagation of particles through the ATLAS detector is modelled using \GEANT~\cite{Agostinelli:2002hh,simulation}. The effect of multiple proton-proton collisions from the same or different bunch crossings is incorporated into the simulation by overlaying additional minimum bias events onto hard-scatter events using \PYTHIA.  Simulated events are weighted to match the distribution of the   number of interactions per bunch crossing observed in data (pile-up).

\section{Event Reconstruction and Preselection}

The data sample was collected with an inclusive selection of  single-lepton and double-lepton triggers.  If the event is selected by the single-lepton triggers, 
 at least one reconstructed muon (electron) is requested to have transverse momentum $p_{\rm T}^{\mu}$ (transverse energy $E_{\rm T}^e$) above 20$\,$GeV (25$\,$GeV).  For di-lepton triggers, at least two leptons are required to be present in the event with transverse energy or momentum above threshold. 
The two muons are required to have $p_{\rm T}^{\mu}$$>$12$\,$GeV for di-muon triggers, and the two electrons to have   $E_{\rm T}^{e}$$>$17$\,$GeV for di-electron triggers, while the thresholds for electron-muon triggers are  $E_{\rm T}^e$$>$15$\,$GeV and  $p_{\rm T}^{\mu}$$>$10$\,$GeV. These thresholds on the reconstructed transverse momenta of leptons are higher than those applied by the online trigger selection, and are chosen such that the trigger efficiency is high, typically between 90 and 99\%, and independent of the transverse momentum of the triggerable objects within uncertainties.

Events recorded during normal running conditions are analysed if the primary vertex has five or more tracks associated to it. The primary vertex of an event is identified as the vertex with the highest $\Sigma p_{\rm T}^2$ of associated tracks.

Electrons must satisfy  ``tight" identification criteria~\cite{electronPerf} and fulfill $|\eta|\,$$<\,$2.47 and $E_{\rm T}\,$$>\,$10$\,$GeV, where $E_{\rm T}$ and $|\eta|$   are determined from the calibrated clustered energy deposits in the electromagnetic calorimeter and the matched ID track respectively. 
Muons are reconstructed by combining tracks in the ID and tracks in the muon spectrometer~\cite{muon}. Reconstructed muons are considered as candidates if they have transverse momentum $p_{\rm T}\,$$>\,$10$\,$GeV and $|\eta|\,$$<\,$2.4.

In this analysis ``tagged" leptons are defined for evaluating the
background, as described below in Sec.~\ref{sec:bkg}. 
Tagged leptons are leptons  separated from each other and from candidate jets as described below. 
If two candidate electrons are reconstructed with $\Delta R \equiv \sqrt{(\Delta\phi)^2+(\Delta\eta)^2}\,$$<\,$0.1, the lower energy one is discarded.  Candidate jets within $\Delta R\,$$=\,$0.2 of an electron candidate are rejected. To suppress leptons originating from  semi-leptonic decays of $c$- and $b$-quarks, all lepton candidates are required to be separated from candidate jets by $\Delta R>0.4$. 
Muons undergoing bremsstrahlung can  be reconstructed with an overlapping electron. To reject these, tagged electrons and muons separated from jets and reconstructed within $\Delta R\,$$=\,$0.1 of each other are both discarded. 
Events containing one or more tagged muons that have transverse impact parameter with respect to the primary vertex $|d_0|\,$$>\,$0.2$\,$mm or longitudinal impact parameter with respect to the primary vertex $|z_0|\,$$>\,$1$\,$mm are rejected to suppress cosmic muon background. 
  
``Signal leptons'' are tagged leptons for which the scalar sum of the transverse momenta of tracks within a cone of $\Delta R\,$$\equiv\,$ $\sqrt{(\Delta\phi)^2+(\Delta\eta)^2}\,$$=\,$0.2 around the lepton candidate, and excluding the lepton candidate track itself, is less than 10\% of the lepton $E_{\rm T}$ for electrons and less than 1.8$\,$GeV for muons.
Tracks selected for the electron and muon isolation requirement, defined above, have $p_{\rm T}\,$$>\,$1$\,$GeV and are associated to the primary vertex of the event. To suppress leptons originating from secondary vertices, the distance of closest approach of the lepton track to the primary vertex normalised to its uncertainty is required to be small, with $|d_{0}|/\sigma(d_{0})\,<\,6(3)$ for electrons (muons).

Jets are reconstructed using the  anti-$k_t$ algorithm~\cite{Cacciari:2008gp} with a radius parameter of $R=0.4$ using clustered energy deposits calibrated at the electromagnetic scale. The jet energy is corrected to account for pile-up and for the non-compensating nature of the calorimeter using correction factors parameterised as a function of the jet $E_{\rm T}$ and $\eta$~\cite{fakeCleanUp}. The correction factors applied to jets have been obtained from simulation and have been tuned and validated using data.
Jets considered in this analysis have $E_{\rm T}\,$$>\,$20$\,$GeV, $|\eta|\,$$<\,$2.5 and a fraction of the jet's track transverse momenta that can be associated with the primary vertex greater than 0.75.  Events containing jets failing the quality criteria described in Ref.~\cite{fakeCleanUp} are rejected to suppress both SM and beam-induced background.  
Jets are identified as containing $b$-hadron decays, and thus called ``$b$-tagged'', using a multivariate technique based on quantities such as the impact parameters of the tracks associated to a reconstructed secondary vertex.  
The $b$-tagging algorithm~\cite{btag} correctly identifies $b$-quark jets in simulated top-quark decays with an efficiency of 60\% and misidentifies jets containing light-flavour quarks and gluons with a rate of $<\,$1\%, for jets with $|\eta|\,$$<\,$2.5 and jet $E_{\rm T}\,$$>\,$20$\,$GeV.

The missing transverse momentum, $\met$, is the magnitude of the vector sum of the transverse momentum or transverse energy of all $p_{\rm T}\,$$>\,$10$\,$GeV muons, $E_{\rm T}\,$$>\,$20$\,$GeV electrons, $E_{\rm T}\,$$>\,$20$\,$GeV jets, and calibrated calorimeter clusters with $|\eta|\,$$<\,$4.9 not associated to these objects~\cite{metPerf}.

\section{Signal Region Selection}

Selected events must contain exactly three signal leptons.
As R-parity conserving leptonic decays of $\ninoi$ yield same-flavour opposite-sign (SFOS) lepton pairs, the presence of at least one such pair is required. The invariant mass of any SFOS lepton pair must be above 20$\,$GeV to suppress background from low-mass resonances and the missing transverse momentum must satisfy $\met\,$$>\,$75$\,$GeV.

Three signal regions are then defined: two ``$Z$-depleted'' regions (SR1a and SR1b), with no SFOS pairs having invariant mass within 10$\,$GeV of the nominal $Z$-boson mass; and a ``$Z$-enriched'' one (SR2), where at least one SFOS pair has an invariant mass within 10$\,$GeV of the $Z$-boson mass. 
Events in SR1a and SR1b are further required to contain no $b$-tagged jets to suppress contributions from $b$-jet-rich background processes, where a  lepton could originate from the decay of a heavy-flavor quark. SR1b is designed to increase sensitivity to scenarios characterised by large mass splittings between the heavy gauginos and the LSP  by requiring all three leptons to have $p_{\rm T}\,>\,$30$\,$GeV.  In both SR1b and SR2, the transverse mass variable $m_{\rm T}$ must take values greater than   90 GeV, where  $m_{\rm T}$ is constructed using the \met  and the lepton not included in the lepton pair with invariant mass closest to the nominal $Z$-boson mass.
The $m_{\rm T}$ requirement is introduced to suppress background from $WZ$ events.
The SR1a/b regions target neutralino decays via intermediate sleptons or via off-shell $Z$ bosons while SR2 targets decays via an on-shell $Z$ boson. 
Table~\ref{tab:SRdef} summarises the selection requirements for the three signal regions.

\begin{table}[t]
\begin{center}
 \caption{The selection requirements for the three signal regions. The $Z$-veto ($Z$-requirement) rejects (selects) events with $m_{\rm SFOS}$ within 10$\,$GeV of the $Z$ mass (91.2$\,$GeV). The $m_{\rm T}$ is calculated from the $E_{\rm T}^{\rm miss}$ and the lepton not forming the best $Z$ candidate. \label{tab:SRdef}}
\footnotesize{
 \begin{tabular}{p{3cm} |c|c|c}

  \hline\hline
  Selection		& SR1a & SR1b & SR2 \\ 
\hline
Targeted Intermediate Decay & \multicolumn{2}{|c|}{$\tilde{l}^{(*)}$ or $Z^{*}$} & on-shell $Z$ \\
\hline
N leptons ($e$, $\mu$) & \multicolumn{3}{|c}{ Exactly 3 } \\
Lepton charge, flavour & \multicolumn{3}{|c}{ At least one SFOS pair with $m_{\ell\ell}>$$\,$20$\,$GeV} \\
$E_{\rm T}^{\rm miss}$ & \multicolumn{3}{|c}{ $>\,$75$\,$GeV }\\
\hline
$m_{\rm SFOS}$ & $Z$-veto & $Z$-veto & $Z$-requirement \\
N $b$-jets & 0 & 0 & any \\
$m_{\rm T}$ & any & $>\,$90$\,$GeV & $>\,$90$\,$GeV \\
$p_{\rm T}\,$ $all \ell$ & $>\,$10$\,$GeV & $>\,$30$\,$GeV & $>\,$10$\,$GeV \\
\hline
\hline
 \end{tabular}
}
 \end{center}
\end{table}

\section{Standard Model Background Estimation}
\subsection{Reducible Background Processes\label{sec:bkg}}
Several SM processes contribute to the background in the signal regions. A ``reducible" process has at least one ``fake" object, that is either a lepton from a semileptonic decay of a heavy-flavour quark or an electron from an isolated photon conversion. The contribution from misidentified light-flavour quark or gluon jets is negligible in the signal regions. The reducible background includes single- and pair-production of top-quarks and $WW$ or $W/Z$ produced in association with jets or photons. The dominant component is the production of top quarks, with a contribution of 1\% or less from $Z$+jets. The reducible background is estimated using a ``matrix method'' similar to that described in Ref.~\cite{mm}. 

In this implementation of the matrix method, the signal lepton with the highest $p_{\rm T}$ or $E_{\rm T}$ is taken to be real, which is a valid assumption in 99\% of the cases, based on simulation. The number of observed events with one or two fakes is then extracted from a system of linear equations relating the number of events with two additional signal or tagged candidates to the number of events with two additional candidates that are either real or fake. 
The coefficients of the linear equations are functions of the real-lepton identification efficiencies and of the fake-object misidentification probabilities. 

The identification efficiency is measured in data using lepton candidates from 
$Z\rightarrow\ell\ell$ decays.
Misidentification probabilities for each relevant fake type (heavy flavour or conversion) and for each reducible background process,  parameterised with the lepton $p_{\rm T}$ and $\eta$, are obtained using simulated events with one signal and two tagged leptons. These misidentification probabilities are then corrected using the ratio (fake scale factor) of the misidentification probability in data to that in simulation obtained from dedicated control samples. For heavy-flavour fakes, the correction factor is measured in a $b\bar{b}$-dominated control sample. This is defined by selecting events with only one $b$-tagged jet (containing a muon) and a tagged lepton, for which the fake rate is measured. 
The non-$b\bar{b}$ background includes top-quark pair production and $W$ bosons produced in association with a $b$-quark. An $\met$ requirement of less than 40$\,$GeV suppresses both the $\ttbar$ and the $W$ contamination, while requiring $m_{\rm T}\,$$<\,$40$\,$GeV reduces the $W$ background. The remaining (small) background is subtracted from data using MC predictions. The fake scale factor for the conversion candidates is  determined in a sample of photons radiated from a muon in $Z\rightarrow\mu\mu$ decays. These are selected by requiring $m_{\mu\mu e}$ to lie within 10$\,$GeV of the nominal $Z$-boson mass value. A weighted average misidentification probability is then calculated by weighting the corrected type- and process-dependent misidentification probabilities according to the relative contributions in a given signal or validation region, defined below. 

\subsection{Irreducible Background Processes}
A background process is considered ``irreducible'' if it leads to events with three real and isolated leptons, referred to as ``real'' leptons below. Such processes include diboson ($WZ$ and $ZZ$) and $t\bar{t}W/Z$ production, where the gauge boson may be produced off-mass-shell. The $ZZ$ and $t\bar{t}W/Z$ contribution is determined using the corresponding MC samples, for which lepton and jet selection efficiencies are corrected to account for differences with respect to data. 

The largest irreducible background, $WZ$, is determined using a semi-data-driven approach. The $WZ$ background is fit to data in a control region  including events with exactly three leptons, one SFOS lepton pair, a $Z$ candidate, $\met\,$$<\,$50$\,$GeV, a $b$-veto, and $m_{\rm T}\,$$>\,$40$\,$GeV. The $WZ$ purity  in the control region  is $\sim$80\%. Non-$WZ$ backgrounds, both irreducible and reducible, are determined based on simulation or by using the matrix method and subtracted. 
A $WZ$ normalisation factor 1.25$\pm$0.12 is obtained in the control region under a background-only hypothesis and used to estimate the $WZ$ background  in the validation regions. To obtain the model-independent 95\% CL upper limit on the new phenomena cross-section, a fit is performed simultaneously in the $WZ$ control region and in the signal region, with floating $WZ$ normalisation factor and a non-negative  signal in the signal region only. This allows the propagation of the uncertainties on the normalisation factor. When setting limits on specific new physics scenarios, the potential signal contamination in the $WZ$  control region is accounted for in the simultaneous fit.

\section{Background Model Validation}
The background predictions have been tested in various validation regions. A region (VR1) dominated by Drell-Yan and $WZ$ events  is selected
by requiring three signal leptons, at least one SFOS lepton pair, 30$\,$GeV$<\,$$\met\,$$<\,$75$\,$GeV, and  a $Z$-boson veto. 
A reducible-background dominated region (VR2, where top-quark pair-production and decay to two real and one fake lepton is the main contribution) is built by requiring three signal leptons,  $\met\,$$>\,$50$\,$GeV and by vetoing SFOS lepton pairs. 
Finally, a $WZ$-dominated region (VR3) is defined by selecting events with three signal leptons, at least one  SFOS lepton pair, a $Z$ candidate, and 50$\,$GeV$<\,$$\met\,$$<\,$75$\,$GeV. The data and predictions are in agreement within the quoted statistical and systematic uncertainties as shown in Table~\ref{tab:VRs}.  

\begin{table}[t]
\begin{center}
 \caption{Expected numbers of events from SM backgrounds  and observed numbers of events in data, for 4.7~fb$^{-1}$, in validation regions VR1, VR2 and VR3. Both statistical and systematic uncertainties are included. \label{tab:VRs}}
\footnotesize{
 \begin{tabular}{l r@{$\pm$}l r@{$\pm$}l r@{}l}

  \hline\hline
  Selection		&\multicolumn{2}{c}{VR1}&\multicolumn{2}{c}{VR2}&\multicolumn{2}{c}{VR3} \\ 
    \hline
    $\ttbar Z$  & 0.17&0.14   & 0.12&0.10   & 1.1&$\pm$0.9   \\
    $\ttbar W$  & 0.6&0.5     & 0.7&0.5   & 0.10&$\pm$0.08   \\
    $\ttbar WW$ & 0.017&0.014 & 0.022&0.017   & 0.0023&$\pm$0.0019   \\
    $ZZ$	     & 17&15   & 0.10&0.05   & 3.9&$\pm$0.6\\
    $WZ$	     & 46&8   & 0.93&0.29   & 98&$\pm$12   \\
    Reducible Bkg.   & 50&28   & 13&7   & 3.1&$^{+4.7}_{-3.1}$   \\
    Total Bkg.	     & 114&32   & 15&7   & 106&$\pm$13   \\ \hline
    Data	     &\multicolumn{2}{c}{126}  &\multicolumn{2}{c}{18}  &\multicolumn{2}{c}{109}  \\
   \hline\hline
 \end{tabular}
}
 \end{center}
\end{table}

\section{Systematic uncertainties}
Several sources of systematic uncertainty are considered in the signal, control and validation regions. 
The systematic uncertainties affecting the simulation-based estimates (the yield of the irreducible background, the cross-section weighted misidentification probabilities, the signal yield) include the theoretical cross-section uncertainties due to renormalisation and factorisation scale and PDFs, the acceptance uncertainty due to PDFs, the uncertainty on the luminosity, the uncertainty due to the jet energy scale, jet energy resolution, lepton energy scale, lepton energy resolution, lepton efficiency, $b$-tagging efficiency, mistag probability, and the choice of MC generator. 
In SR1a, the total uncertainty on the irreducible background is 24\%. This is dominated by
 the uncertainty on the efficiency of the signal region selection for the $WZ$  generator, determined by comparing the nominal yield with that obtained with the \Herwig generator and found to be 20\%. The next largest uncertainties are the uncertainty
 due to the MC generator  (16\%)  and that 
    on the cross-sections (9\%)  of the non-$WZ$ background.  The MC generator uncertainty partially accounts for the cross-section uncertainty, leading to a slight overestimate of the overall uncertainty. 
 All the remaining uncertainties on the irreducible background in this signal region range between 0.5 and 5\%. The total uncertainty on the irreducible background in SR1b is slightly larger, at 25\%, due to the limited number of simulated events. In SR2, the uncertainty on the irreducible background is 24\%, with increased contributions from the jet energy scale and resolution and cross-section uncertainties. 

The uncertainty on the reducible background includes the MC uncertainty on the weights for the misidentification probabilities from the sources listed above (up to 10\%) and the uncertainty due to the dependence of the misidentification probability on $\met$ (0.6--15\%). Also included in the uncertainty on the reducible background is the uncertainty on the fake scale factors (10--34\%), and that due to the limited number of data events with three tagged leptons, of which at least one is a signal lepton (19--130\%). The latter uncertainty is highest in SR2 where the reducible background is very low.

The total uncertainties on the signal yields are  10--20\%, where the largest contribution is from the uncertainty on the cross-sections (7\%). Signal cross-sections are calculated to NLO in the strong coupling constant using \Prospino. An envelope of cross-section predictions is defined using the 68\% CL ranges of the \CTEQSix~\cite{Nadolsky:2008zw} (including the $\alpha_S$ uncertainty) and the  \MSTW~\cite{Martin:2009iq} PDF sets, together with variations of the factorisation and renormalisation scales by factors of two or one half. The nominal cross-section value is taken to be the midpoint of the envelope and the uncertainty assigned is half the full width of the envelope,  following the PDF4LHC recommendations~\cite{Botje:2011sn}.

In all of the above, the value used for the uncertainty on the luminosity is 3.9\%~\cite{lumipaper,lumiconf}.
Correlations of systematic uncertainties between processes and regions are accounted for.

\section{Results and Interpretation}
\begin{figure}[ht]
\centering
\subfigure[]{  \includegraphics[width=8cm]{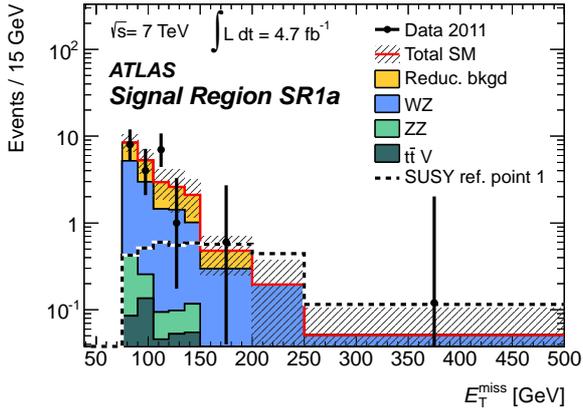}}
 \subfigure[]{  \includegraphics[width=8cm]{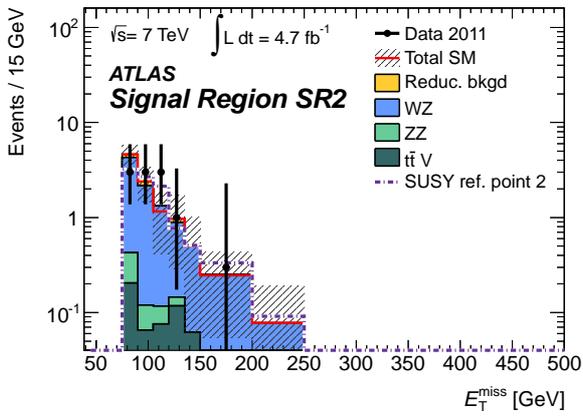}}
\caption{\label{fig:met} $\met$ distributions for events in signal regions SR1a (a) and SR2 (b). 
The uncertainty band includes both statistical and systematic uncertainty, while the uncertainties on the data points are statistical only.
The yields for two of the simplified model scenarios are also shown for illustration purposes: one with intermediate sleptons ``SUSY ref. point 1''  ($m_{\chinoone}, m_{\ninotwo} , m_{\tilde{\ell}_L}, m_{\tilde{\chi}_1^{0}}\,$$=\,$425, 425, 250, 75$\,$GeV) and a second with no sleptons ``SUSY ref. point 2'' ($m_{\chinoone}, m_{\ninotwo}, m_{\tilde{\chi}_1^{0}}\,$$=\,$150, 150, 0$\,$GeV). The signal distribution is not stacked on top of the expected background. 
}
\end{figure}

The numbers of observed events and the prediction for SM backgrounds in SR1a, SR1b and SR2 are given in Table~\ref{tab:SRs}. 
 Distributions of the $\met$ in SR1a and SR2 are presented in Fig.~\ref{fig:met}.

\begin{table}[t]
\begin{center}
 \caption{Expected numbers of events from SM backgrounds and observed numbers of events in data, for 4.7~fb$^{-1}$, in signal regions SR1a, SR1b and SR2. 
The yield for two of the simplified model scenarios, ``SUSY ref. point 1'' with intermediate sleptons, ($m_{\chinoone}, m_{\ninotwo}, m_{\tilde{\ell}_L}, m_{\tilde{\chi}_1^{0}}\,$$=\,$425, 425, 250, 75$\,$GeV) and ``SUSY ref. point 2'' with no intermediate sleptons, ($m_{\chinoone}, m_{\ninotwo} , m_{\tilde{\chi}_1^{0}}\,$$=\,$150, 150, 0$\,$GeV) are also presented.  Both statistical and systematic uncertainties are included. Upper limits on the observed and expected visible production cross-section at 95\% CL are also shown. \label{tab:SRs}}
\footnotesize{
 \begin{tabular}{l r@{$\pm$}l r@{$\pm$}l r@{}l}
  \hline\hline
  Selection		&\multicolumn{2}{c}{SR1a}&\multicolumn{2}{c}{SR1b}&\multicolumn{2}{c}{SR2} \\ 
    \hline
  \midrule
    $\ttbar Z$ & 0.06&0.05 & 0.025&0.023   & 0.6&$\pm$0.5 \\ 
    $\ttbar W$ & 0.36&0.29 & 0.10&0.08   & 0.09&$\pm$0.08 \\ 
    $\ttbar WW$& 0.010&0.008 & 0.0023&0.0019   & 0.004&$\pm$0.004 \\ 
    $ZZ$	   & 0.67&0.21 & 0.09&0.08   & 0.34&$\pm$0.17 \\ 
    $WZ$	   & 13.5&2.9 & 1.05&0.28   & 9.3&$\pm$2.1 \\ 
    Reducible Bkg. & 10&5 & 0.35&0.34   & 0.5&$^{+1.0}_{-0.5}$ \\ 
    Total Bkg.	   & 25&6 & 1.6&0.5   & 10.9&$\pm$2.4 \\ \hline
    Data	     &\multicolumn{2}{c}{24}   &\multicolumn{2}{c}{0}  &\multicolumn{2}{c}{11} \\
   \hline  \hline
    SUSY ref. point 1 & 8.0 & 0.8 & 6.5 & 0.6 & 0.46&$\pm$0.05 \\
    SUSY ref. point 2 & 1.03 & 0.19 & 0.21 & 0.09 & 10.9&$\pm$1.0 \\
 \hline  
Visible $\sigma$ (exp) & \multicolumn{2}{c}{$<\,$3.0$\,$fb} & \multicolumn{2}{c}{$<\,$0.8$\,$fb} & \multicolumn{2}{c}{$<\,$2.0$\,$fb} \\
Visible $\sigma$ (obs) & \multicolumn{2}{c}{$<\,$3.0$\,$fb} & \multicolumn{2}{c}{$<\,$0.7$\,$fb} & \multicolumn{2}{c}{$<\,$2.0$\,$fb} \\

   \hline\hline
 \end{tabular}
}
 \end{center}
\end{table}

No significant excess of events is found in any of the three signal regions. 
Upper limits on the visible cross-section, defined as the production cross-section times acceptance times efficiency, of 3.0$\,$fb in SR1a, 0.7$\,$fb in SR1b and 2.0$\,$fb in SR2 are placed at 95\% CL with the modified frequentist {\it CL$_s$} prescription~\cite{CLs}. All systematic uncertainties and their correlations are taken into account via nuisance parameters in a profile likelihood fit~\cite{cowan}.  
 The corresponding expected limits are 3.0$\,$fb, 0.8$\,$fb and 2.0$\,$fb, respectively. 

\begin{figure}[tp]
 \begin{center}
  \begin{picture}(250,540)

   \put(30,400){\epsfig{figure=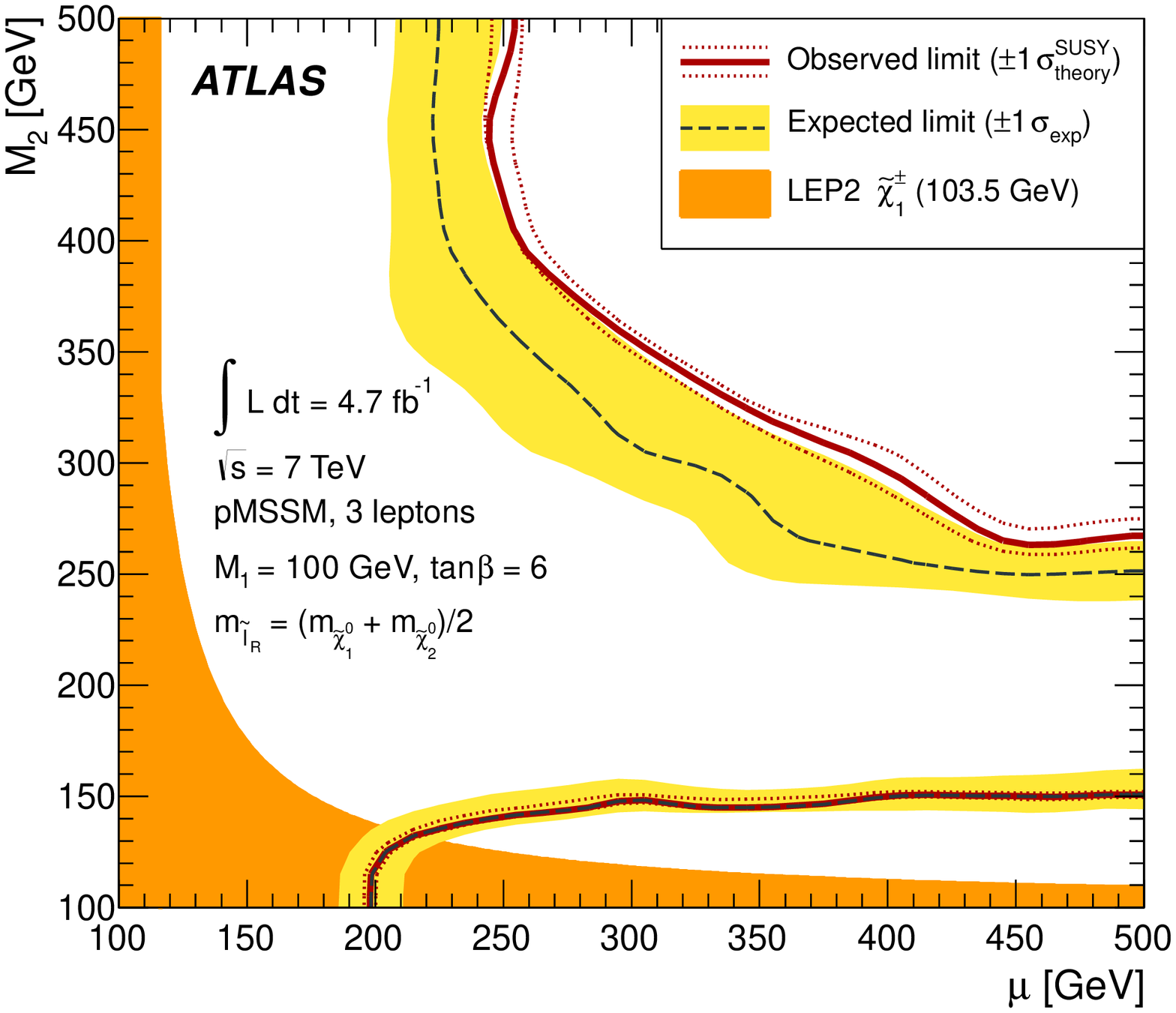,width=7.5cm}}
   \put(120,398){\small{(a)}} 

   \put(30,200){\epsfig{figure=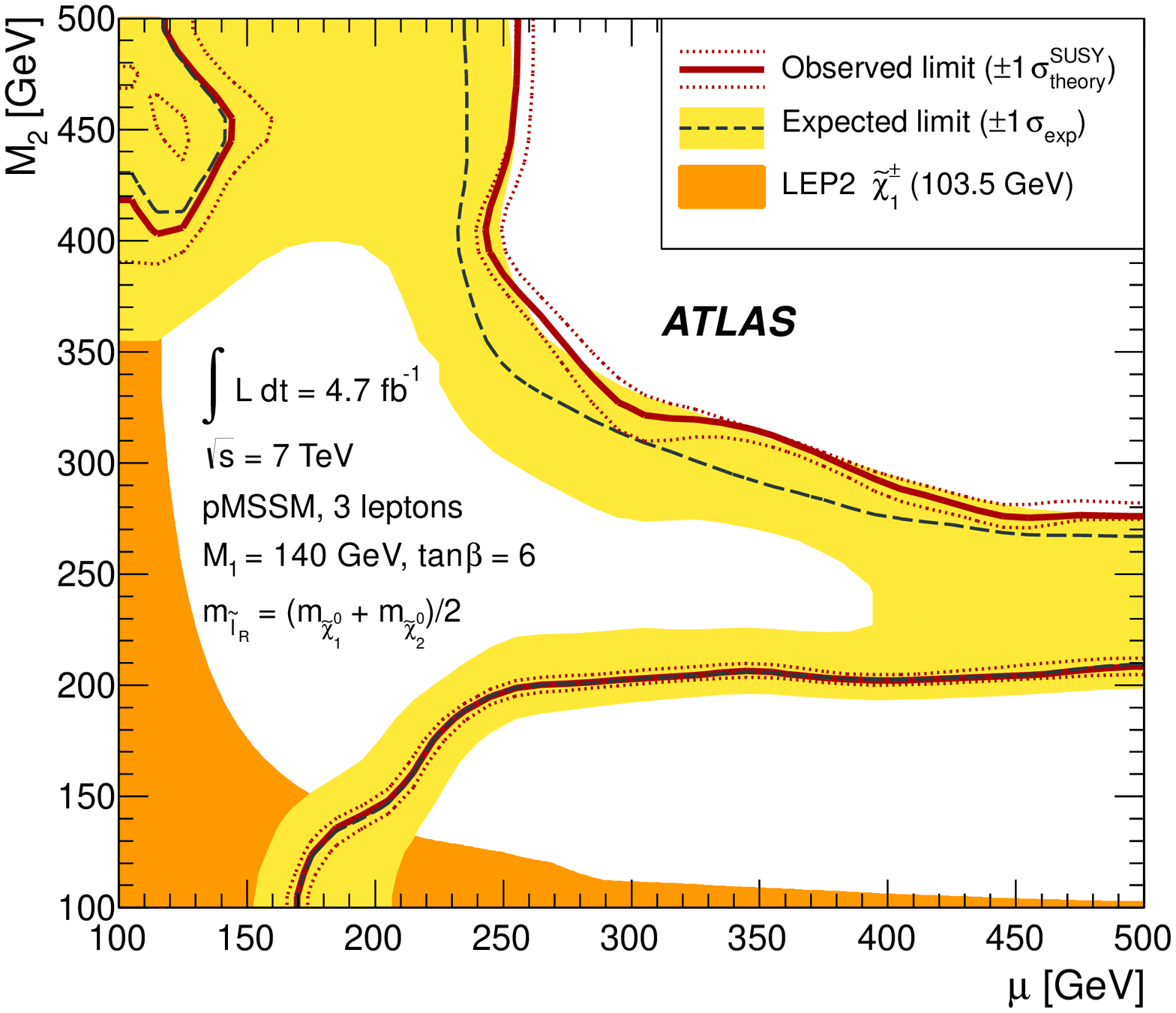,width=7.5cm}}
   \put(120,198){\small{(b)}} 

   \put(30,0){\epsfig{figure=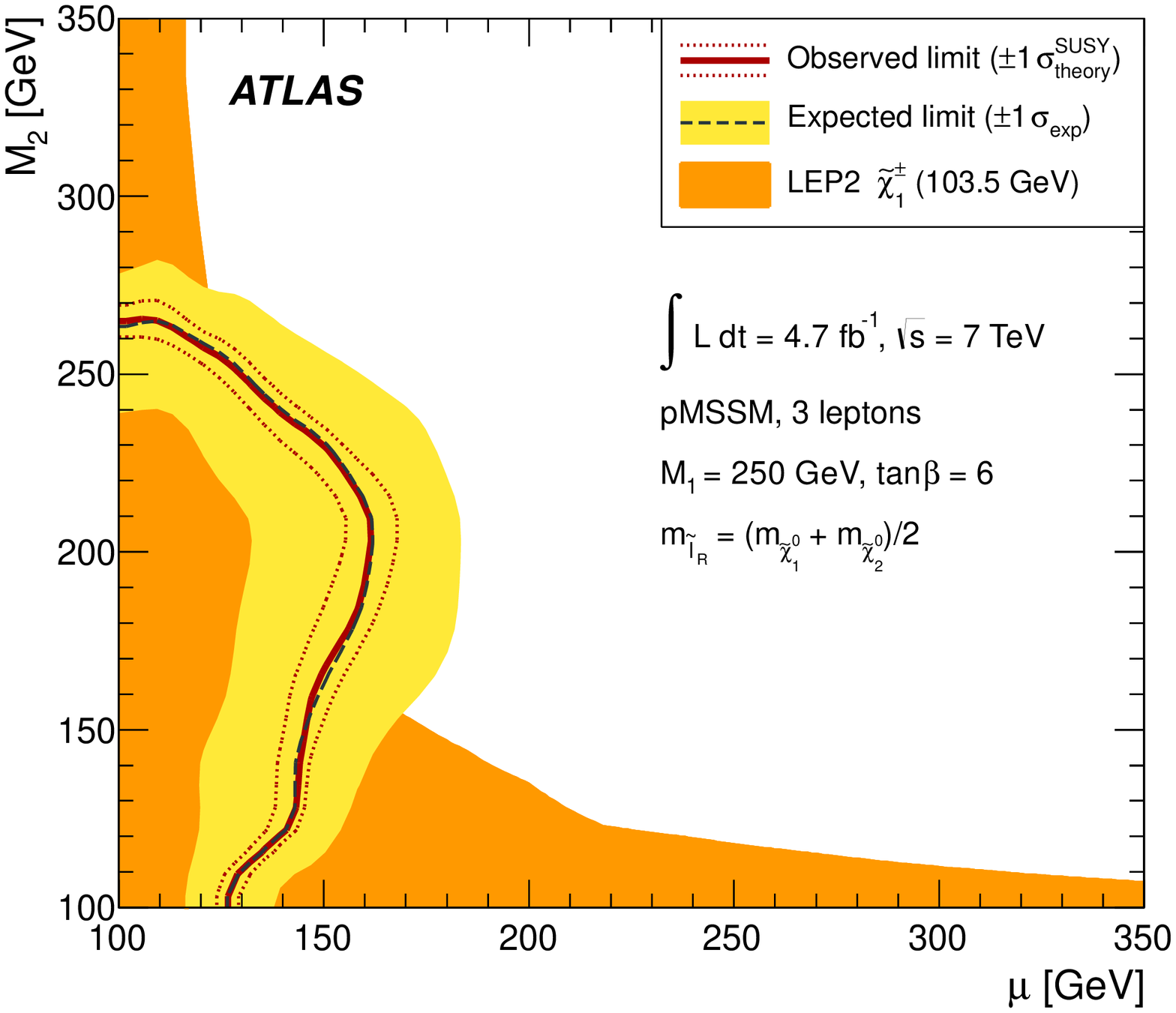,width=7.5cm}}
   \put(120,0){\small{(c)}}

  \end{picture}

 \caption{Observed and expected 95\% CL limit contours for chargino and neutralino production in the pMSSM for $M_{1}\,$$=\,$100$\,$GeV (a), $M_{1}\,$$=\,$140$\,$GeV (b) and $M_{1}\,$$=\,$250$\,$GeV (c). The regions with low values of $M_2$ and $\mu$ are the excluded ones for all values of $M_{1}$. 
  The expected and observed limits are calculated without signal cross-section uncertainty taken into account. The yellow band is the $\pm$1$\sigma$ experimental uncertainty on the expected limit (black dashed line). The red dotted band is the $\pm$1$\sigma$ signal theory uncertainty on the observed limit (red solid line). 
 The LEP2 limit in the Figure corresponds to the limit on the $\chinoone$ mass in\protect{~\cite{PDG}} as transposed to this pMSSM plane. 
 Linear interpolation is used to account for the discreteness of the signal grids. The exclusion contours are optimised by applying in each signal grid point the CL values from the most sensitive signal region (lowest expected CL) for $M_{1}\,$$=\,$100$\,$GeV and 140$\,$GeV, whereas signal region SR1a is used for $M_{1}\,$$=\,$250$\,$GeV.  \label{fig:pMSSMlimits}}
 \end{center}
\end{figure}

SR1a and SR1b provide the best sensitivity for the pMSSM scenarios; in particular SR1a (SR1b)  targets scenarios with small (large) mass splitting between the heavy gauginos and the LSP. 
The limits are calculated using the signal region providing the best expected limit for each of the model points.
The uncertainties on the signal cross-section are not included in the limit calculation but their impact on the observed limit is shown. 
The exclusion limits for the pMSSM are shown in Fig.~\ref{fig:pMSSMlimits}  as a function of the three parameters $M_1$ , $M_2$ and
$\mu$, where the regions with low values of $M_2$ and $\mu$ are the excluded ones for all values of $M_1$. In these plots, the main features can be explained in broad terms as follows. For a given value of $M_{1}\,$, for example  $M_{1}\,$$=\,$100$\,$GeV in Fig.~\ref{fig:pMSSMlimits} (a), the production cross-section decreases as $M_{2}\,$ and $\mu$ increase, which explains why limits become less stringent when both  $M_{2}\,$ and $\mu$ take high values. In general, the sensitivity is reduced in the region at low  $M_{2}\,$ and high $\mu$, due to the small mass splitting between the \ninotwo and the \ninoone. When  $\mu$ is greater than  $M_{1}\,$ and $M_{2}\,$, which is true for example in the rightmost part of the exclusion plots for $M_{1}\,$$=\,$100$\,$GeV (Fig.~\ref{fig:pMSSMlimits} (a)) and $M_{1}\,$$=\,$140$\,$GeV  (Fig.~\ref{fig:pMSSMlimits}      (b)), the mass of the gauginos does not depend on $\mu$ and the sensitivity remains constant as a function of $\mu$. On the contrary, in a large section of the plane shown for $M_{1}\,$$=\,$250$\,$GeV  (Fig.~\ref{fig:pMSSMlimits} (c)), the condition that $\mu$  should be greater than $M_{1}\,$ is not fulfilled and the resulting limits on the same plane become less stringent. Additionally, the reduced reach at high $M_{2}\,$ and low $\mu$  for $M_{1}\,$$=\,$140 GeV can be explained in terms of smaller cross-section values and smaller mass splittings in that section of the parameter space. The 
difference between expected and observed limits seen in the upper right corner of the $M_{1}\,$$=\,$100 GeV exclusion plot, where SR1b has the best sensitivity, is explained by the observed under-fluctuation in data with respect to SM predictions. The value of $\tan\beta$ does not have a significant
impact on $\sigma(pp\rightarrow\chinojpm\ninoi)\times BR(\chinojpm\ninoi\rightarrow\ell\nu\tilde{\chi}_1^{0}\ell\ell\tilde{\chi}_1^{0})$, which decreases  by 10\% if $\tan\beta$ is raised from 6 to 10. 

\begin{figure}[tp]
 \begin{center}
  \begin{picture}(250,540)

   \put(30,430){\epsfig{figure=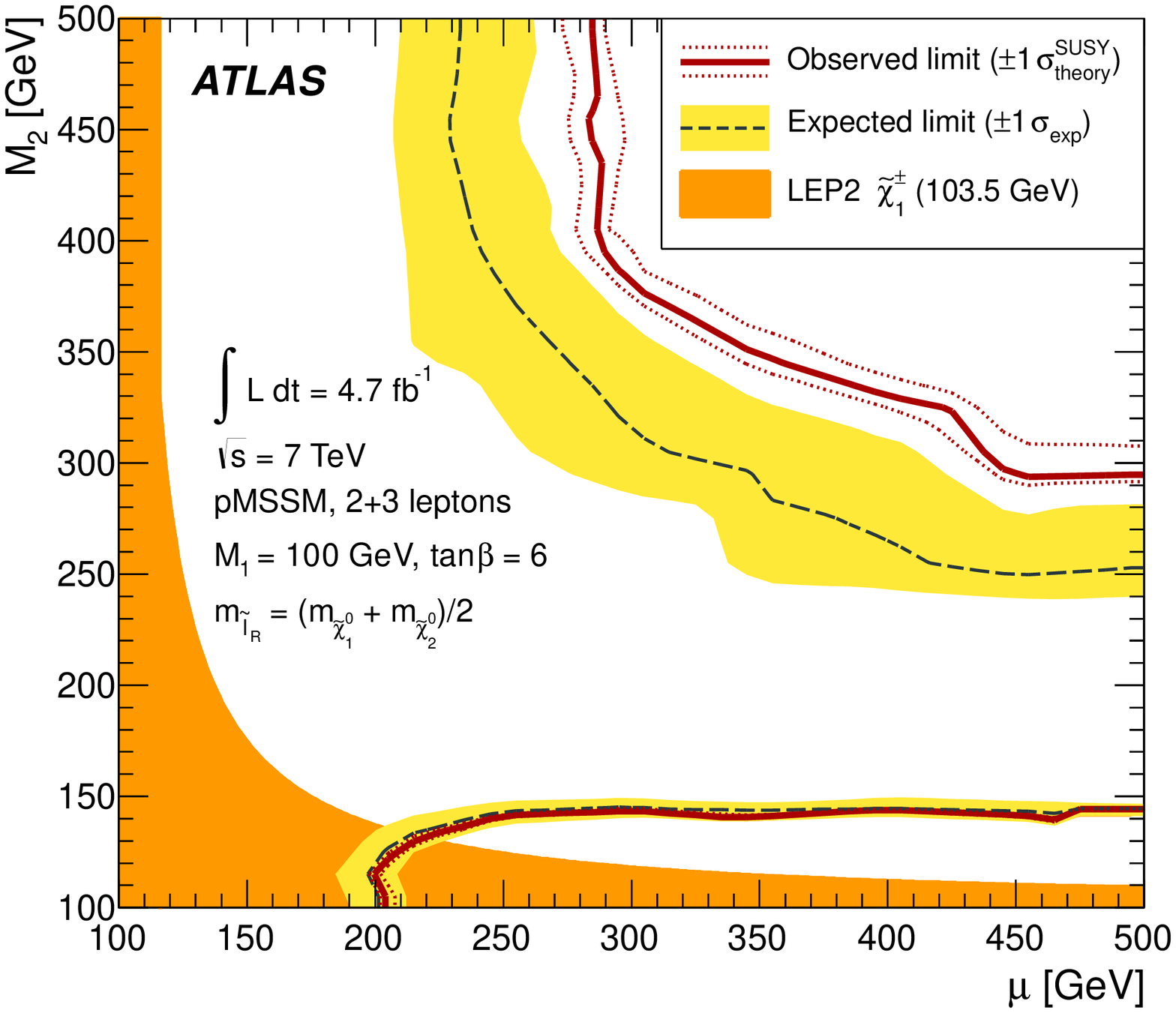,width=7.5cm}}
   \put(120,428){\small{(a)}} 

   \put(30,232){\epsfig{figure=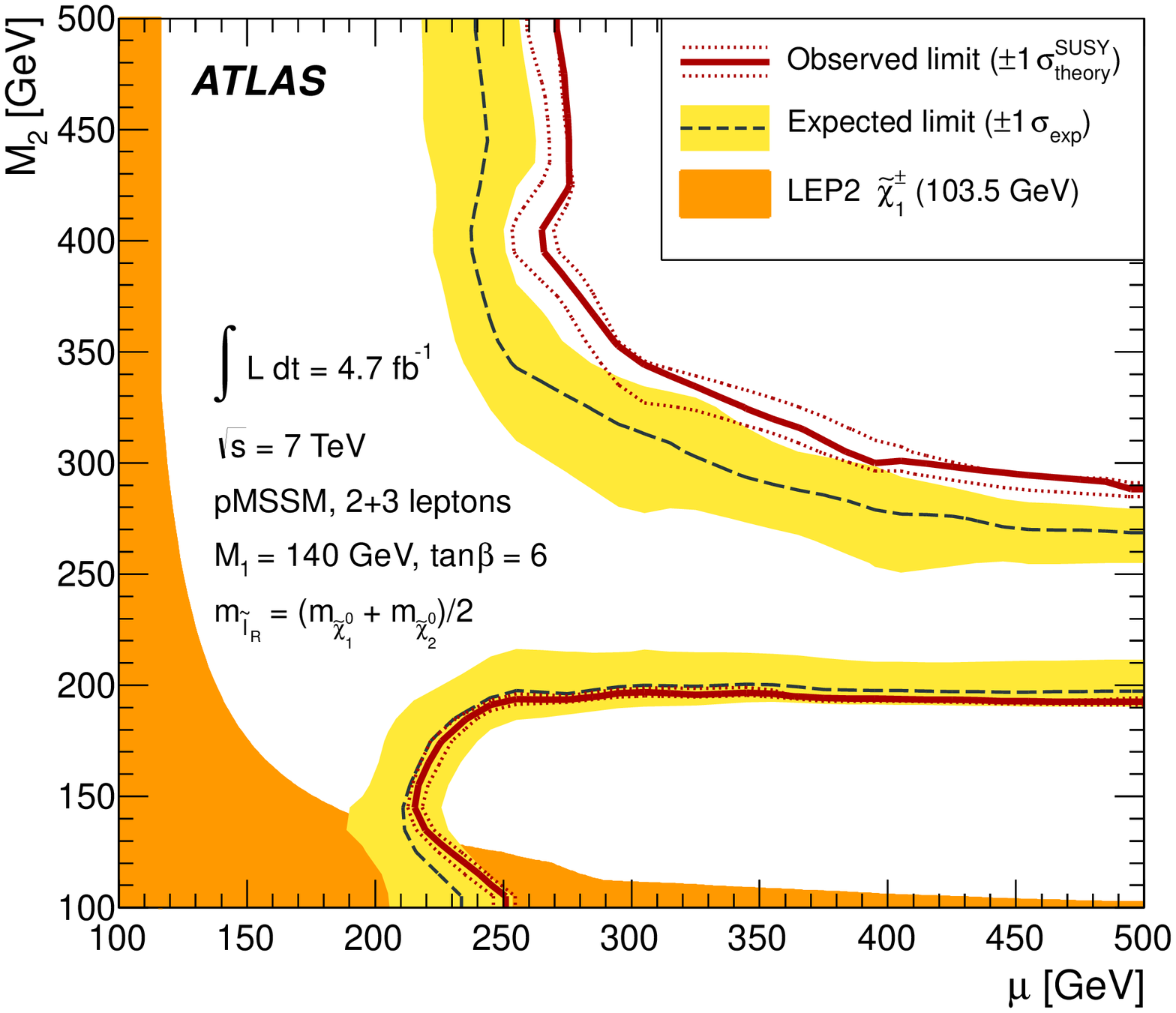,width=7.5cm}}
   \put(120,230){\small{(b)}} 

   \put(30,30){\epsfig{figure=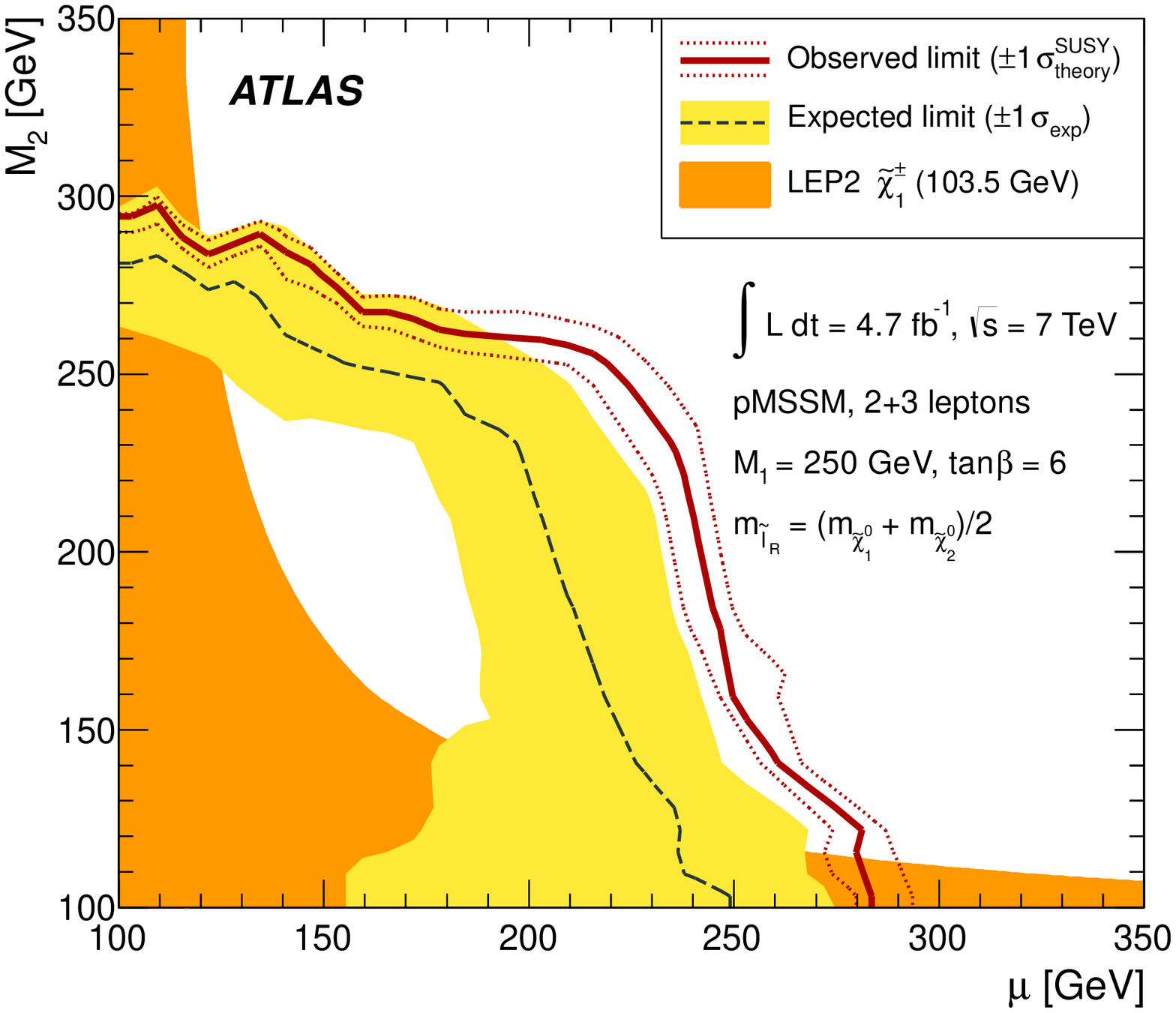,width=7.5cm}}
   \put(120,28){\small{(c)}}
  \end{picture}
 \caption{Observed and expected 95\% CL limit contours for chargino and neutralino production in the pMSSM for $M_{1}\,$$=\,$100$\,$GeV (a), $M_{1}\,$$=\,$140$\,$GeV (b) and $M_{1}\,$$=\,$250$\,$GeV (c). 
 Contours from the combination of the results from this search with those of
the two-lepton ATLAS search in~\cite{2L7TeV}. The various limits are as described in Figure~\ref{fig:pMSSMlimits}.
 The colour coding is the same as that in Figure~\ref{fig:pMSSMlimits}. \label{fig:pMSSMlimits2L3L}}
\end{center}
\end{figure}

The results obtained in signal regions SR1a and SR1b are combined with results  
from the relevant signal region  in the ATLAS two-lepton search (SR-m$_{\rm T2}$)~\cite{2L7TeV}. The fits are performed on the combined likelihood function from SR-m$_{\rm T2}$ with SR1a,  and from SR-m$_{\rm T2}$ with SR1b.
 The combination yielding the highest expected sensitivity is selected	for optimal exclusions in the pMSSM planes (Fig.~\ref{fig:pMSSMlimits2L3L}). The uncertainties are profiled in the likelihood and correlations between channels and processes are taken into account. An improvement in the sensitivity  for $M_{1}\,$$=\,$250$\,$GeV and small values of $M_{2}\,$ is seen
 when results from the three-lepton and the two-lepton analyses are combined.

\begin{figure}[t]
 \subfigure[]{ \includegraphics[width=8.5cm]{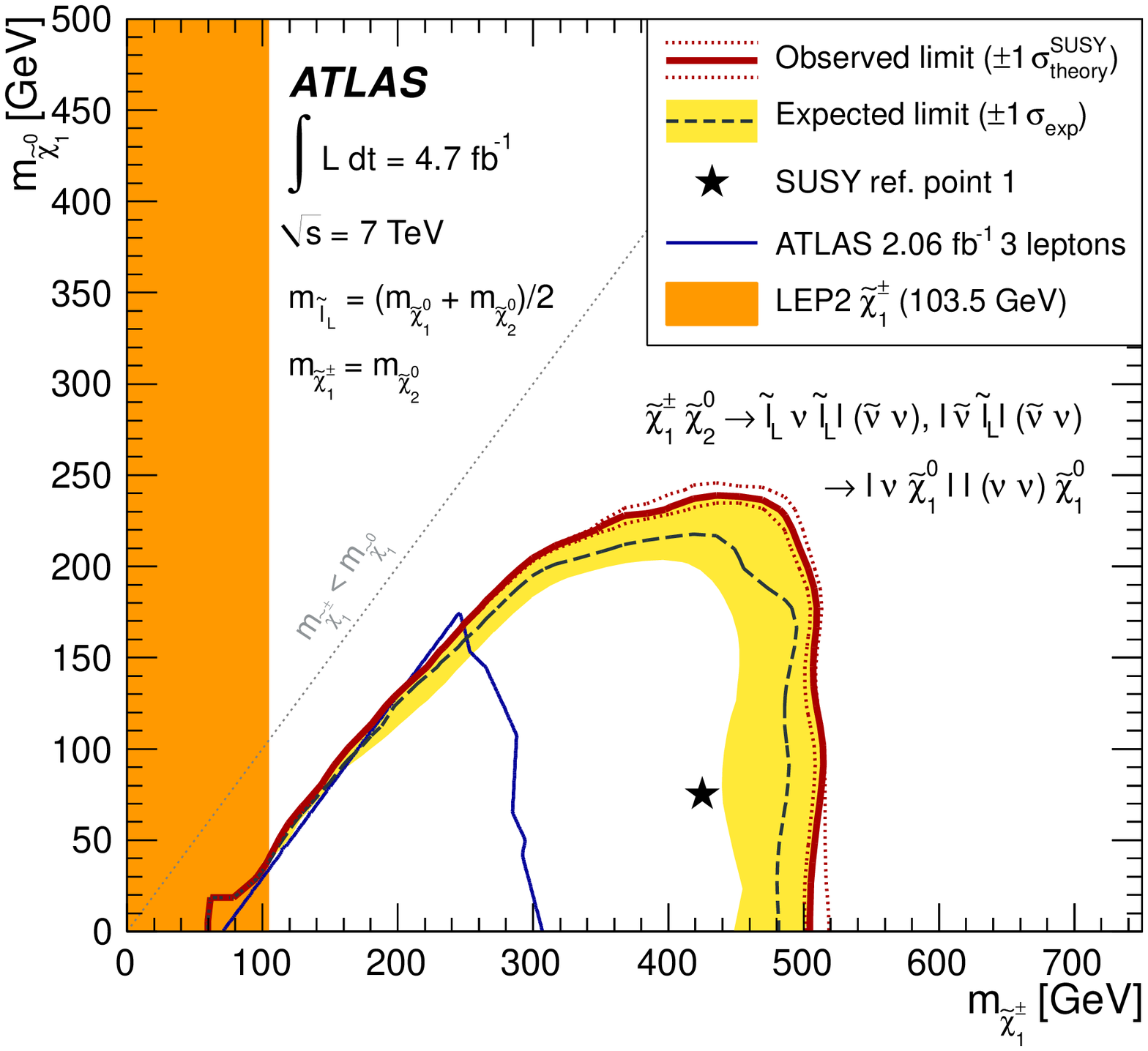}}
\subfigure[]{ \includegraphics[width=8.5cm]{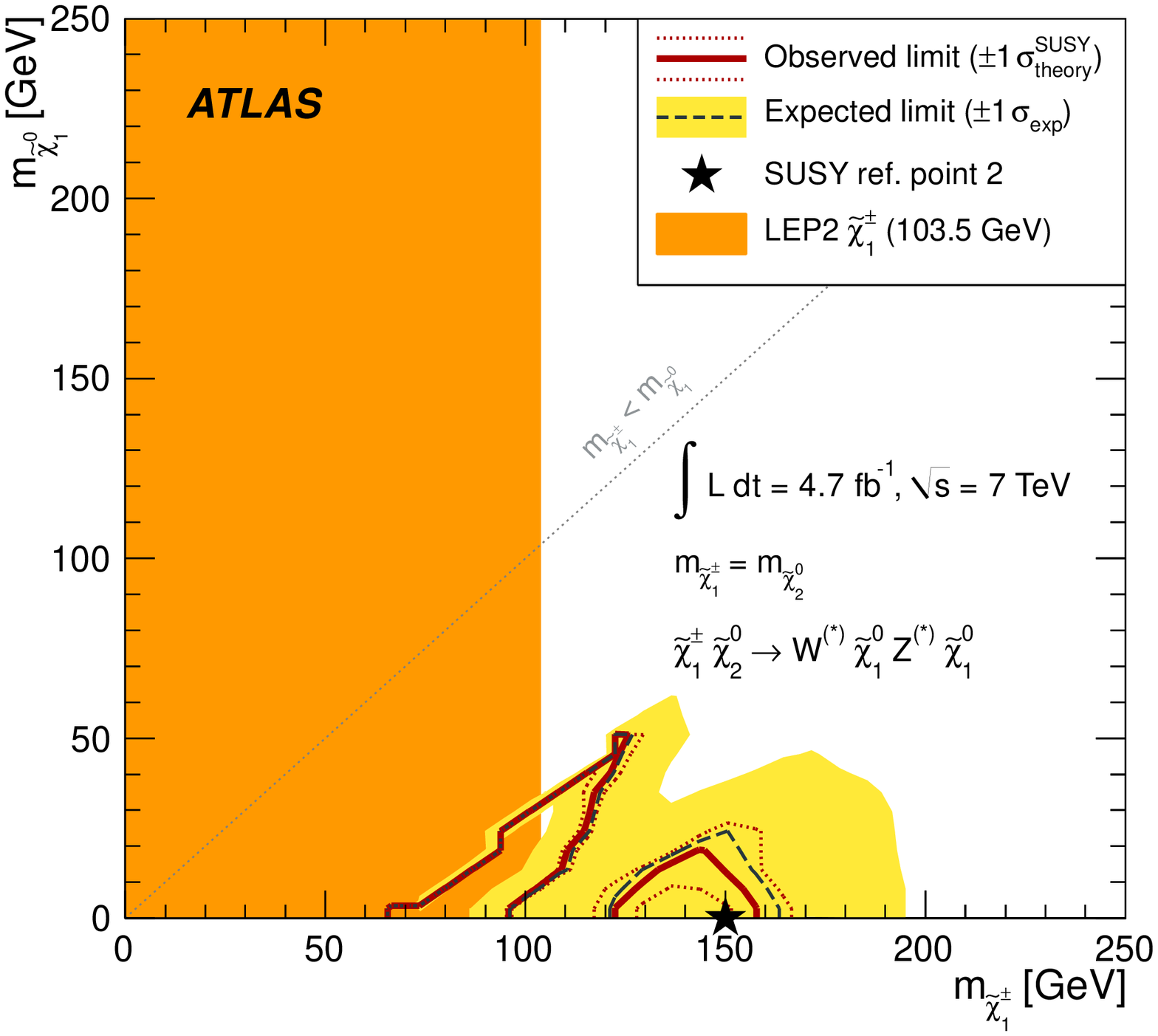}}
 \caption{Observed and expected 95\% CL limit contours for chargino and neutralino production in the simplified model scenario with intermediate slepton decay (a) and intermediate gauge boson decay (b). The colour coding is the same as that in  Figure~\ref{fig:pMSSMlimits}. For  scenarios with intermediate slepton decay (with no intermediate slepton decay) the reference point is  ``SUSY ref. point 1" (``SUSY ref. point 2").  The ``ATLAS 2.06 fb$^{-1}$ 3 leptons" contour corresponds to the result of the ATLAS search documented in~\cite{ATLASRel163LSUSY}.  \label{fig:Simplimits}}

\end{figure}

 Region SR1b provides the best sensitivity to the  simplified models with intermediate slepton decay for which the interpretation is shown in Fig.~\ref{fig:Simplimits} (a). In the simplified models with intermediate slepton decays, degenerate $\chinoone$ and $\ninotwo$ masses up to 500$\,$GeV are excluded for large mass differences from the $\tilde{\chi}_1^{0}$. Both SR1a and SR2 are used to interpret the results in the simplified model with gauginos decaying via gauge bosons (Fig.~\ref{fig:Simplimits} (b)). The signal region SR1a has the best sensitivity for small mass differences between the heavy and light neutralinos, while SR2 is sensitive to decays of \ninotwo\  into on-mass-shell $Z$ bosons. 

\section{Summary}
Results from a search for direct production of charginos and neutralinos in the final state with three leptons (electrons or muons) and missing transverse momentum are reported. The analysis is based on 
4.7$\,$fb$^{-1}$ of proton-proton collision data delivered by the LHC at $\sqrt{s}=$7$\,$TeV and collected by ATLAS. 
No significant excess of events is found in data. The null result is interpreted in the pMSSM and simplified models. 
For the pMSSM, an improvement in the sensitivity  for $M_{1}\,$$=\,$250$\,$GeV and small values of $M_{2}\,$ is seen when results from this analysis are combined with those from the corresponding two-lepton ATLAS search. 
For the simplified models  with intermediate slepton decays, degenerate $\chinoone$ and $\ninotwo$ masses up to 500$\,$GeV are excluded for large mass differences from the $\tilde{\chi}_1^{0}$. The analysis presented   
here also has sensitivity to direct gaugino production with decays via   
gauge bosons.

\section{Acknowledgements}

We thank CERN for the very successful operation of the LHC, as well as the
support staff from our institutions without whom ATLAS could not be
operated efficiently.

We acknowledge the support of ANPCyT, Argentina; YerPhI, Armenia; ARC,
Australia; BMWF, Austria; ANAS, Azerbaijan; SSTC, Belarus; CNPq and FAPESP,
Brazil; NSERC, NRC and CFI, Canada; CERN; CONICYT, Chile; CAS, MOST and NSFC,
China; COLCIENCIAS, Colombia; MSMT CR, MPO CR and VSC CR, Czech Republic;
DNRF, DNSRC and Lundbeck Foundation, Denmark; EPLANET and ERC, European Union;
IN2P3-CNRS, CEA-DSM/IRFU, France; GNAS, Georgia; BMBF, DFG, HGF, MPG and AvH
Foundation, Germany; GSRT, Greece; ISF, MINERVA, GIF, DIP and Benoziyo Center,
Israel; INFN, Italy; MEXT and JSPS, Japan; CNRST, Morocco; FOM and NWO,
Netherlands; RCN, Norway; MNiSW, Poland; GRICES and FCT, Portugal; MERYS
(MECTS), Romania; MES of Russia and ROSATOM, Russian Federation; JINR; MSTD,
Serbia; MSSR, Slovakia; ARRS and MVZT, Slovenia; DST/NRF, South Africa;
MICINN, Spain; SRC and Wallenberg Foundation, Sweden; SER, SNSF and Cantons of
Bern and Geneva, Switzerland; NSC, Taiwan; TAEK, Turkey; STFC, the Royal
Society and Leverhulme Trust, United Kingdom; DOE and NSF, United States of
America.

The crucial computing support from all WLCG partners is acknowledged
gratefully, in particular from CERN and the ATLAS Tier-1 facilities at
TRIUMF (Canada), NDGF (Denmark, Norway, Sweden), CC-IN2P3 (France),
KIT/GridKA (Germany), INFN-CNAF (Italy), NL-T1 (Netherlands), PIC (Spain),
ASGC (Taiwan), RAL (UK) and BNL (USA) and in the Tier-2 facilities
worldwide.

\bibliographystyle{elsarticle-num}
\raggedright
\bibliography{directgaugino3L_7TeV}

\clearpage 
\onecolumn

\begin{flushleft}
{\Large The ATLAS Collaboration}

\bigskip

G.~Aad$^{\rm 48}$,
T.~Abajyan$^{\rm 21}$,
B.~Abbott$^{\rm 111}$,
J.~Abdallah$^{\rm 12}$,
S.~Abdel~Khalek$^{\rm 115}$,
A.A.~Abdelalim$^{\rm 49}$,
O.~Abdinov$^{\rm 11}$,
R.~Aben$^{\rm 105}$,
B.~Abi$^{\rm 112}$,
M.~Abolins$^{\rm 88}$
% ...
\bigskip
\\
% ...
$^{11}$ Institute of Physics, Azerbaijan Academy of Sciences, Baku, Azerbaijan\\
$^{12}$ Institut de F{\'\i}sica d'Altes Energies and Departament de F{\'\i}sica de la Universitat Aut{\`o}noma de Barcelona and ICREA, Barcelona, Spain\\
% ...
$^{21}$ Physikalisches Institut, University of Bonn, Bonn, Germany\\
% ...
$^{48}$ Fakult{\"a}t f{\"u}r Mathematik und Physik, Albert-Ludwigs-Universit{\"a}t, Freiburg, Germany\\
$^{49}$ Section de Physique, Universit{\'e} de Gen{\`e}ve, Geneva, Switzerland\\
% ...
$^{88}$ Department of Physics and Astronomy, Michigan State University, East Lansing MI, United States of America\\
% ...
$^{105}$ Nikhef National Institute for Subatomic Physics and University of Amsterdam, Amsterdam, Netherlands\\
% ...
$^{111}$ Homer L. Dodge Department of Physics and Astronomy, University of Oklahoma, Norman OK, United States of America\\
$^{112}$ Department of Physics, Oklahoma State University, Stillwater OK, United States of America\\
% ...
$^{115}$ LAL, Universit{\'e} Paris-Sud and CNRS/IN2P3, Orsay, France\\
% ...
\end{flushleft}


\end{document}